# On the Feasibility of Network Alignment for Three-Source Three-Destination Multiple Unicast Networks with Delays

Abhinav Ganesan, Teja Damodaram Bavirisetti and B. Sundar Rajan
Email: {abhig_88, bsrajan}@ece.iisc.ernet.in, dbaviris@broadcom.com

*Abstract*—A transform approach to network coding was introduced by Bavirisetti et al. (arXiv:1103.3882v3 [cs.IT]) as a tool to view wireline networks with delays as $k$-instantaneous networks (for some large $k$). When the local encoding kernels (LEKs) of the network are varied with every time block of length $k > 1$, the network is said to use block time varying LEKs. In this work, we propose a Precoding Based Network Alignment (PBNA) scheme based on transform approach and block time varying LEKs for three-source three-destination multiple unicast network with delays (3-S 3-D MUN-D). In a recent work, Meng et al. (arXiv:1202.3405v1 [cs.IT]) reduced the infinite set of sufficient conditions for feasibility of PBNA in a three-source three-destination instantaneous multiple unicast network as given by Das et al. (arXiv:1008.0235v1 [cs.IT]) to a finite set and also showed that the conditions are necessary. We show that the conditions of Meng et al. are also necessary and sufficient conditions for feasibility of PBNA based on transform approach and block time varying LEKs for 3-S 3-D MUN-D.

## I. INTRODUCTION

The notion of Network Coding was introduced in [1] where the capacity of wireline multicast networks is characterized. Scalar linear network coding was found to achieve the capacity of multicast networks [2]. In the meanwhile, it was shown that [3] there exist solvable non-multicast networks where scalar linear network coding is insufficient. In addition, [3] also showed that determining the existence of linear network coding solution for multiple unicast networks is NP-hard in general. In [4], it was conjectured that vector linear network coding suffices to solve networks with arbitrary message demands. Subsequently, Dogherty et al. [5] disproved the conjecture by showing that there exists networks where vector linear network coding does not achieve network capacity and that nonlinear network coding are required in general. However, the practicality of linear network codes led to construction of suboptimal network codes for Multiple Unicast networks based on linear programming [6].

The concept of interference alignment originally introduced in interference channels [7] was applied by Das et al. [8], [9] in a three-source three-destination instantaneous multiple unicast network (3-S 3-D I-MUN), where the zero interference conditions of Koetter et al. [10] cannot be met, to achieve a rate of half for each source destination pair. Since precoding matrices are used at the sources for interference alignment and exploited for network coding in 3-S 3-D I-MUN, it came to be known as Precoding Based Network Alignment (PBNA). Though PBNA is not optimal in general for a 3-S 3-D I-MUN [9], it provides a simple and systematic manner of network code construction that can guarantee (under certain conditions) an asymptotic rate of half for every source destination pair when the zero interference conditions cannot be met.

Sufficient conditions for feasibility of PBNA in a 3-S 3-D I-MUN were obtained in [8]. However, the set of conditions were infinite and hence, impossible to check. Moreover, the sufficient conditions were constrained by the use of particular precoding matrices at the sources. These motivated the work of Meng et al. [11] where, a finite set of conditions are obtained for feasibility of PBNA in a 3-S 3-D I-MUN that are both necessary and sufficient. We call these finite set of conditions as the "reduced feasibility conditions". The highlight of their result is that PBNA with arbitrary precoding matrices is feasible iff PBNA is feasible with the choice of precoding matrices as in [8] (with the number of symbol extensions being greater than or equal to five). The derivation of the result involved taking into account graph related properties.

A Discrete Fourier Transform (DFT) based approach to acyclic networks with delays was introduced by Bavirisetti et al. [12] for arbitrary acyclic networks with delays. The primary result of the work is that acyclic networks with delays can be transformed into $k$ instantaneous networks (for some large $k$). This transform approach enabled the application of PBNA in three-source three-destination multiple unicast network with delays (3-S 3-D MUN-D) to achieve a throughput of half for every source destination pair, where the zero interference conditions cannot be met. It was also shown that, unlike in 3-S 3-D I-MUN, there exists 3-S 3-D MUN-D where PBNA based on time-invariant local encoding kernels (LEKs) is feasible. The PBNA was then generalized with the use of time-varying LEKs and algebraic necessary and sufficient conditions for feasibility of PBNA in 3-S 3-D MUN-D were obtained. However, these conditions are applicable to only to the case of precoding over a fixed number of symbol extensions, i.e., if the feasibility test fails over a symbol extension of length $k$, it is not known if the test would fail for a symbol extension of length greater than $k$. Hence, in the absence of an elegant set of conditions that would help check the feasibility of PBNA in a 3-S 3-D MUN-D over any number of symbol extensions (like in [11]), we are motivated to look for an alternative PBNA scheme for 3-S 3-D MUN-D. In this work, we shall propose a PBNA scheme and show that its feasibility conditions inherit the reduced feasibility conditions of Meng et al.

*Definition 1:* A 3-S 3-D MUN-D is said to use block time varying LEKs when the LEKs are varied with every time block of length $k > 1$ and remain constant within each time block.

The contributions of the paper are as follows:

- A PBNA scheme for 3-S 3-D MUN-D based on transform approach and block time varying LEKs is proposed.
- Necessary and sufficient conditions for feasibility of the proposed PBNA scheme is shown to be the same as the reduced feasibility conditions for 3-S 3-D I-MUN.

The organization of this paper is as follows. In Section II, we shall briefly review the system model and the transform approach to 3-S 3-D MUN-D. The PBNA scheme for 3-S 3-D MUN-D based on transform approach and block time varying LEKs shall be detailed in Section III. The necessary and sufficient conditions for feasibility of the proposed scheme will be discussed in Section IV. Section V will conclude the paper.

*Notations:* For a variable $p$ which takes integer values between 0 to $k-1$ where $k$ is a positive integer, the notation $A^{(p)}$ denotes matrices indexed by $p$. The notation $diag(A_1, A_2, \cdots, A_n)$ represents a diagonal matrix whose diagonal elements are $A_1, A_2, \cdots, A_n$.

## II. BACKGROUND

In this section, we shall review the system model for 3-S 3-D MUN-D [10] and decomposition of 3-S 3-D MUN-D into $k$ instantaneous networks [12].

A 3-S 3-D MUN-D is a network where Source-$i$, denoted by $S_i$, needs to communicate with Destination-$i$, denoted by $T_i$ ($i \in \{1, 2, 3\}$). The min-cut between $S_i$ and $T_i$ is assumed to be 1. We consider a 3-S 3-D MUN-D represented by a Directed Acyclic Graph (DAG) $\mathcal{G} = (V, E)$, where $V$ is the set of nodes and $E$ is the set of directed links. Arbitrary (positive) integer delay on each link is assumed. We assume that every directed link between a pair of nodes represents an error-free link and has a capacity of one $\mathbb{F}_{2^m}$ symbol per link-use for some positive integer $m > 0$.

The input random processes $X_i(D)$ of $S_i$, output random processes $Y_i(D)$ at $T_i$ and random processes $Z_e(D)$ transmitted on the link $e$ are considered as a power series in a delay parameter $D$, i.e., $X_i(D) = \sum_{t=0}^{\infty} X_i^{(t)} D^t$, $Y_i(D) = \sum_{t=0}^{\infty} Y_i^{(t)} D^t$ and $Z_e(D) = \sum_{t=0}^{\infty} Z_e^{(t)} D^t$ where, $X_i^{(t)}$, $Y_i^{(t)}$ and $Z_e^{(t)}$ denote the input symbol of $S_i$, output symbol of $T_i$ and the symbol transmitted on link $e$ respectively at time instant $t$.

Scalar linear network coding is assumed on the 3-S 3-D MUN-D. The symbol transmitted on a link $e$ at time instant $(t+1)$ is given by

$$Z_e^{(t+1)} = \sum_{i=1}^{3} \alpha_{i,e} X_i^{(t)} + \sum_{e': \text{head}(e') = \text{tail}(e)} \beta_{e',e} Z_{e'}^{(t)}$$

where, $(\alpha_{i,e}, \beta_{e',e}) \in \mathbb{F}_{2^m}$ and $\alpha_{i,e} = 0$ when $S_i \neq \text{tail}(e)$, for all $i$. The output symbol of $T_i$ at time instant $(t+1)$ is given by $Y_i^{(t+1)} = \sum_{e': \text{head}(e') = T_i} \epsilon_{e',i} Z_{e'}^{(t)}$ where, $\epsilon_{e',i} \in \mathbb{F}_{2^m}$. The scalars $\alpha_{j,e}, \beta_{e',e}$ and $\epsilon_{e',j}$ are called local encoding kernels (LEKs) denoted by $\underline{\varepsilon}$. The output random process at $T_j$ can be written in terms of the transfer matrix from $S_i$ to $T_j$, given by $M_{ij}(D)$, as [10]

$$Y_j(D) = \sum_{i=1}^{3} M_{ij}(D) X_i(D) \quad (1)$$

where, $M_{ij}(D) = \sum_{d=0}^{d_{max}} M_{ij}^{(d)} D^d$ where, $d_{max}$ is the difference between the maximum and minimum of the path delays from $S_i$ to $T_j$, over all $(i, j)$, between which a path exists. Note that here, $M_{ij}(D)$ is a $1 \times 1$ matrix. $M_{ij}(D)$ is also a function of the LEKs and is explicitly indicated only when required (i.e., denoted by $M_{ij}(\underline{\varepsilon}, D)$).

When the LEKs are varied with time, denote the set of LEKs from time instant $t_1$ to time instant $t_2$ ($t_2 \geq t_1$) by $\underline{\varepsilon}^{(t_1, t_2)}$, i.e., $\underline{\varepsilon}^{(t_1, t_2)} = \{\underline{\varepsilon}^{(t_1)}, \underline{\varepsilon}^{(t_1+1)}, \ldots, \underline{\varepsilon}^{(t_2)}\}$ where, $\underline{\varepsilon}^{(t_i)}$ denotes the LEKs at time $t_i$. The output symbols of $T_j$ at time instant $t$ is given by [12]

$$Y_j^{(t)} = \sum_{i=1}^{s} \sum_{d=0}^{d_{max}} M_{ij}^{(d)}(\underline{\varepsilon}^{(t-d,t)}) X_i^{(t-d)}. \quad (2)$$

When the LEKs are time invariant, $M_{ij}^{(d)}(\underline{\varepsilon}^{(t-d,t)}) = M_{ij}^{(d)}$. The details of the exact dependence of $M_{ij}^{(d)}(\underline{\varepsilon}^{(t-d,t)})$ on $\underline{\varepsilon}^{(t-d,t)}$ can be seen in [12]. We note that the output symbol at time $t$ at any destination depends only on the LEKs $\underline{\varepsilon}^{(t-d_{max}, t)}$.

### A. Review of the Transform Approach of [12]

Denote a $k$-length input symbol of $S_i$ by $X_i^k$, i.e, $X_i^k = \begin{bmatrix} X^{(k-1)} & X^{(k-2)} & \cdots & X^{(0)} \end{bmatrix}^T$. Similarly denote a $k$-length output symbol at $T_i$ by $Y_i^k$. The set of $\mathbb{F}_{2^m}$-symbols generated by the sources at any particular time instant are said to constitute the same generation. Consider the transmission scheme where, given $k (>> d_{max})$ generations of input symbols at each source, the last $d_{max}$ generations is transmitted first (which is called the *cyclic prefix*) followed by the $k$ generations of input symbols. In effect, $k + d_{max}$ time slots at each source are used for transmitting $k$ generations. Then, $Y_j^{(k+d_{max})}$ is written as (3) using (1). Then, $Y_j(D)$ can be written as (3) using (1). Discarding the first $d_{max}$ outputs at $T_j$, (3) is re-written as (4). It is assumed that $k$ divides $2^m - 1$.

*Theorem 1 ( [12]):* The matrix $M_{ij}$, as defined in (4), can be diagonalized as $M_{ij} = F \hat{M}_{ij} F^{-1}$ where, $\hat{M}_{ij} = diag\left(\hat{M}_{ij}^{(k-1)}, \hat{M}_{ij}^{(k-2)}, \ldots, \hat{M}_{ij}^{(0)}\right)$. The elements $\hat{M}_{ij}^{(l)}$ ($l \in \{0, 1, \cdots (k-1)\}$) are given by $\hat{M}_{ij}^{(l)} = \sum_{d=0}^{d_{max}} \alpha^{(k-1-l)d} M_{ij}^{(d)}$ and the matrix $F$ is the DFT matrix given by

$$F = \begin{bmatrix} 1 & 1 & 1 & \cdots & 1 \\ 1 & \alpha & \alpha^2 & \cdots & \alpha^{k-1} \\ 1 & \alpha^2 & \alpha^4 & \cdots & \alpha^{2(k-1)} \\ \vdots & \vdots & \vdots & \vdots & \vdots \\ 1 & \alpha^{k-1} & \alpha^{2(k-1)} & \cdots & \alpha^{(k-1)(k-1)} \end{bmatrix}.$$

where, $\alpha \in \mathbb{F}_{2^m}$ and $\alpha^k = 1$.

At each source $S_i$, transmit $X_i'^k = F X_i^k$ instead of $X_i^k$. Then, at each destination $T_j$, the output symbol vector of length $k$ given by $Y_j'^k$ is pre-multiplied by $F^{-1}$ after

$$\begin{bmatrix} Y_j^{(k-1)} \\ Y_j^{(k-2)} \\ \vdots \\ Y_j^{(0)} \\ Y_j^{(-1)} \\ \vdots \\ Y_j^{(-d_{max})} \end{bmatrix} = \sum_{i=1}^{s} \begin{bmatrix} M_{ij}^{(0)} & M_{ij}^{(1)} & \cdots & M_{ij}^{(d_{max})} & 0 & 0 & \cdots & 0 & 0 \\ 0 & M_{ij}^{(0)} & \cdots & M_{ij}^{(d_{max}-1)} & M_{ij}^{(d_{max})} & 0 & \cdots & 0 & 0 \\ \vdots & \vdots & \vdots & \vdots & \vdots & \ddots & \ddots & \vdots & \vdots \\ 0 & 0 & \cdots & 0 & M_{ij}^{(0)} & M_{ij}^{(1)} & \cdots & M_{ij}^{(d_{max}-1)} & M_{ij}^{(d_{max})} \\ 0 & 0 & \cdots & 0 & 0 & M_{ij}^{(0)} & \cdots & M_{ij}^{(d_{max}-2)} & M_{ij}^{(d_{max}-1)} \\ \vdots & \vdots & \vdots & \vdots & \vdots & \vdots & \vdots & \vdots & \vdots \\ 0 & 0 & \cdots & 0 & 0 & 0 & 0 & 0 & M_{ij}^{(0)} \end{bmatrix} \begin{bmatrix} X_i^{(k-1)} \\ X_i^{(k-2)} \\ \vdots \\ X_i^{(0)} \\ X_i^{(k-1)} \\ \vdots \\ X_i^{(k-d_{max})} \end{bmatrix} \quad (3)$$

$$\begin{bmatrix} Y_j^{(k-1)} \\ Y_j^{(k-2)} \\ \vdots \\ Y_j^{(0)} \end{bmatrix} = \sum_{i=1}^{s} \underbrace{\begin{bmatrix} M_{ij}^{(0)} & M_{ij}^{(1)} & \cdots & M_{ij}^{(d_{max}-1)} & M_{ij}^{(d_{max})} & 0 & \cdots & 0 & 0 & 0 \\ 0 & M_{ij}^{(0)} & \cdots & M_{ij}^{(d_{max}-2)} & M_{ij}^{(d_{max}-1)} & M_{ij}^{(d_{max})} & \cdots & 0 & 0 & 0 \\ \vdots & \vdots & \vdots & \vdots & \vdots & \vdots & \vdots & \vdots & \vdots & \vdots \\ M_{ij}^{(1)} & M_{ij}^{(2)} & \cdots & M_{ij}^{(d_{max})} & 0 & 0 & \cdots & 0 & 0 & M_{ij}^{(0)} \end{bmatrix}}_{M_{ij}} \begin{bmatrix} X_i^{(k-1)} \\ X_i^{(k-2)} \\ \vdots \\ X_i^{(0)} \end{bmatrix} \quad (4)$$

discarding the first $d_{max}$ symbols. Therefore, by application of the above theorem, the effective output symbol is given by

$$Y_j^k = F^{-1} Y_j'^k = F^{-1} \sum_{i=1}^{s} F \hat{M}_{ij} F^{-1} F X_i^k = \sum_{i=1}^{s} \hat{M}_{ij} X_i^k \quad (5)$$

The 3-S 3-D MUN-D is now said have been transformed into $k$ instantaneous networks.

## III. PBNA Using Transform Approach and Block Time Variant LEKs

In this section, we propose a PBNA scheme different from that given in [12] for 3-S 3-D MUN-D. Consider the following transmission where, every source $S_i$ is required to transmit a $k(2n+1)$-length block of symbols ($k >> d_{max}$) given by $[X_i^{(1)} \ X_i^{(2)} \cdots \ X_i^{(k(2n+1))}]^T$ for some positive integer $n > 0$. Partition the block of symbols into $(2n+1)$ blocks, each of length $k$ symbols. For each block of $k$ symbols, we add a cyclic prefix of length $d_{max}$. The partitioning of input symbols and addition of cyclic prefix (CP) are shown in Fig. 1.

The LEKs of the network are varied with every $(k+d_{max})$ time instants starting from the time instant $t = -d_{max}$. Therefore, when $S_i$ transmits its first block of data as shown in Fig. 1, the LEKs remain constant and when it starts the transmission of the second block of data, the LEKs encountered in the network are different.

At each destination $T_i$, discard the first $d_{max}$ outputs in each received block of length $(k+d_{max})$ symbols, starting from time instant $t = -d_{max}$. This is termed as discarding the cyclic prefix. Denote the LEKs during $l^{\text{th}}$-block transmission by $\underline{\varepsilon}_l$ ($1 \leq l \leq (2n+1)$). Now, consider the second block of output symbols (i.e., $l = 2$) at $T_j$ after discarding the cyclic prefix. Since the LEKs remain constant during one block of transmission, from (2) and (3), we get (6) (at the top of the next page). As in (4), (6) is re-written as (7). Using Theorem 1, $M_{ij}(\underline{\varepsilon}_2)$ can be diagonalized to $\hat{M}_{ij}(\underline{\varepsilon}_2)$. Similarly, the $l^{\text{th}}$-block of output symbols, after discarding the cyclic prefix, can be written in terms of the matrix $\hat{M}_{ij}(\underline{\varepsilon}_l)$ ($1 \leq l \leq (2n+1)$). We note that

$$\hat{M}_{ij}(\underline{\varepsilon}_l) = diag\left(M_{ij}(\underline{\varepsilon}_l, 1), \ M_{ij}(\underline{\varepsilon}_l, \alpha), \cdots, \ M_{ij}(\underline{\varepsilon}_l, \alpha^{(k-1)})\right) \quad (8)$$

Let $X_1'^{(n+1)k}$, $X_2'^{nk}$, and $X_3'^{nk}$ denote the $(n+1)k$-length, $nk$-length, and $nk$-length independent symbols generated by $S_1$, $S_2$, and $S_3$ respectively. Partition each of the independent input symbols into $k$ blocks. Denote the $p^{\text{th}}$-block of independent input symbols of $S_i$ by $X_i'(p)$ ($0 \leq p \leq k-1$) which is a column vector of lengths $(n+1)$ for $S_1$, $n$ for $S_2$, and $n$ for $S_3$. The symbols $X_i'(p)$ ($0 \leq p \leq k-1$) are precoded onto $X_i^{(2n+1)}$ as follows. Define $X_i^{(p \oplus k)} = \left[X_i^{(p)} \ X_i^{(p+k)} \ X_i^{(p+2k)} \cdots \ X_i^{(p+2nk)}\right]^T$ ($0 \leq p \leq k-1$). Let $V_i^{(p)}$ denote the precoding matrices at $S_i$ ($0 \leq p \leq k-1$). The matrices, for all $p$, are of size $(2n+1) \times (n+1)$, $(2n+1) \times n$ and $(2n+1) \times n$ for $i = 1, 2$, and 3 respectively. Now, the symbols to be transmitted by $S_i$, before pre-multiplication by $F$ and addition of CP, are given by $X_i^{(p \oplus k)} = V_i^{(p)} X_i'(p)$. In brief, the $p^{\text{th}}$ element of every block to be transmitted by $S_i$, before pre-multiplication by $F$ and the addition of CP, are obtained by precoding the $p^{\text{th}}$ block of independent symbols $X_i'(p)$. The instance of $p = 0$ is shown in Fig. 1.

After discarding the CP and pre-multiplying by $F^{-1}$ at $T_j$, we obtain $(2n+1)k$-output symbols. These are partitioned into $k$-blocks, each of length $(2n+1)$-symbols. Each block is given by $Y_i^{(p \oplus k)} = \left[Y_i^{(p)} \ Y_i^{(p+k)} \ Y_i^{(p+2k)} \cdots \ Y_i^{(p+2nk)}\right]^T$ ($0 \leq p \leq k-1$). The input-output relation is now given by

$$Y_i^{(p \oplus k)} = \sum_{i=1}^{3} diag\left(M_{ij}(\underline{\varepsilon}_1, \alpha^p), \ M_{ij}(\underline{\varepsilon}_2, \alpha^p),\right. \quad (9)$$
$$\left. \cdots, \ M_{ij}(\underline{\varepsilon}_{2n}, \alpha^p), \ M_{ij}(\underline{\varepsilon}_{(2n+1)}, \alpha^p)\right) V_i^{(p)} X_i'(p).$$

Let $M_{ij}^p = diag\left(M_{ij}(\underline{\varepsilon}_1, \alpha^p), \ \cdots, \ M_{ij}(\underline{\varepsilon}_{(2n+1)}, \alpha^p)\right)$.

## IV. Feasibility of PBNA using Transform Approach and Block Time Varying LEKs

We assume that the min-cut between $S_i - T_j$ is not zero for all $i \neq j$. The proof technique for feasibility of PBNA in the case of min-cut between $S_i - T_j$ being zero for some $i \neq j$ will be similar to that used for non-zero min-cut.

PBNA using Transform Approach and Block Time Varying LEKs requires that the following conditions be satisfied [8] for $0 \leq p \leq k-1$.

$\text{Span}(M_{31}^p V_3^{(p)}) \subset \text{Span}(M_{21}^p V_2^{(p)}), \text{Span}(M_{32}^p V_3^{(p)}) \subset \text{Span}(M_{12}^p V_1^{(p)}),$
$\text{Span}(M_{23}^p V_2^{(p)}) \subset \text{Span}(M_{13}^p V_1^{(p)}),$

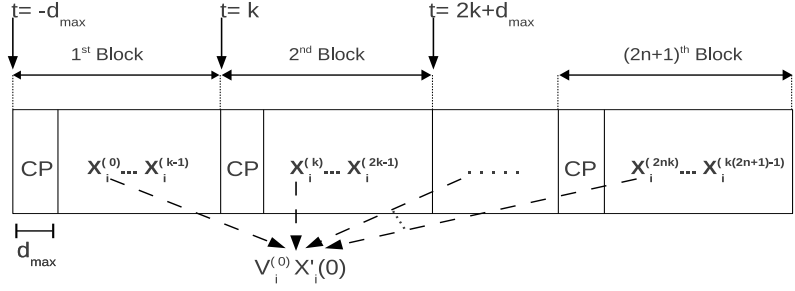

Fig. 1. The figure demonstrates the transmission of $(2n+1)$ blocks of symbols, involving addition of CP for every block at $S_i$. The pre-multiplication of each block of symbols by $F$ (not explicitly shown in the figure) is done after the precoding step and before the addition of CP.

$$\begin{bmatrix}Y_j^{(2k+d_{max}-1)}\\Y_j^{(2k+d_{max}-2)}\\\vdots\\Y_j^{(k+d_{max})}\end{bmatrix}=\sum_{i=1}^{s}\begin{bmatrix}M_{ij}^{(0)}(\underline{\varepsilon}_2)&M_{ij}^{(1)}(\underline{\varepsilon}_2)&\cdots&M_{ij}^{(d_{max})}(\underline{\varepsilon}_2)&0&0&\cdots&0&0\\0&M_{ij}^{(0)}(\underline{\varepsilon}_2)&\cdots&M_{ij}^{(d_{max}-1)}(\underline{\varepsilon}_2)&M_{ij}^{(d_{max})}(\underline{\varepsilon}_2)&0&\cdots&0&0\\\vdots&\vdots&\vdots&\vdots&\ddots&\ddots&\ddots&\vdots&\vdots\\0&0&\cdots&0&M_{ij}^{(0)}(\underline{\varepsilon}_2)&M_{ij}^{(1)}(\underline{\varepsilon}_2)&\cdots&M_{ij}^{(d_{max}-1)}(\underline{\varepsilon}_2)&M_{ij}^{(d_{max})}(\underline{\varepsilon}_2)\end{bmatrix}$$
$$\times\begin{bmatrix}X_i^{(2k-1)}&X_i^{(2k-2)}&\cdots&X_i^{(k)}&X_i^{(2k-1)}&\cdots&X_i^{(2k-d_{max})}\end{bmatrix}^T \quad (6)$$

$$\begin{bmatrix}Y_j^{(2k+d_{max}-1)}\\Y_j^{(2k+d_{max}-2)}\\\vdots\\Y_j^{(k+d_{max})}\end{bmatrix}=\sum_{i=1}^{s}\underbrace{\begin{bmatrix}M_{ij}^{(0)}(\underline{\varepsilon}_2)&M_{ij}^{(1)}(\underline{\varepsilon}_2)&\cdots&M_{ij}^{(d_{max}-1)}(\underline{\varepsilon}_2)&M_{ij}^{(d_{max})}(\underline{\varepsilon}_2)&0&\cdots&0&0&0\\0&M_{ij}^{(0)}(\underline{\varepsilon}_2)&\cdots&M_{ij}^{(d_{max}-2)}(\underline{\varepsilon}_2)&M_{ij}^{(d_{max}-1)}(\underline{\varepsilon}_2)&M_{ij}^{(d_{max})}(\underline{\varepsilon}_2)&\cdots&0&0&0\\\vdots&\vdots&\vdots&\vdots&\vdots&\vdots&\vdots&\vdots&\vdots&\vdots\\M_{ij}^{(1)}(\underline{\varepsilon}_2)&M_{ij}^{(2)}(\underline{\varepsilon}_2)&\cdots&M_{ij}^{(d_{max})}(\underline{\varepsilon}_2)&0&0&\cdots&0&0&M_{ij}^{(0)}(\underline{\varepsilon}_2)\end{bmatrix}}_{M_{ij}(\underline{\varepsilon}_2)}\begin{bmatrix}X_i^{(2k-1)}\\X_i^{(2k-2)}\\\vdots\\X_i^{(k)}\end{bmatrix} \quad (7)$$

$\mathrm{Rank}[M_{11}^p V_1^{(p)} \quad M_{21}^p V_2^{(p)}] = \mathrm{Rank}[V_1^{(p)} \quad M_{11}^{p\,-1} M_{21}^p V_2^{(p)}] = 2n+1$ (10)

$\mathrm{Rank}[M_{22}^p V_2^{(p)} \quad M_{12}^p V_1^{(p)}] = \mathrm{Rank}[M_{12}^{p\,-1} M_{22}^p V_2^{(p)} \quad V_1^{(p)}] = 2n+1$

$\mathrm{Rank}[M_{33}^p V_3^{(p)} \quad M_{13}^p V_1^{(p)}] = \mathrm{Rank}[M_{13}^{p\,-1} M_{33}^p V_3^{(p)} \quad V_1^{(p)}] = 2n+1$

We first note that recovering $X_i'(0)$, for all $i$, represents the feasibility problem of PBNA in the instantaneous version of the original 3-S 3-D MUN-D. Suppose that we cannot recover $X_i'(0)$, for all $i$. But, if we can recover $X_i'(p)$, for all $p \neq 0$ and for all $i$, we can still achieve throughputs of $\frac{(n+1)(k-1)}{(2n+1)k}$, $\frac{n(k-1)}{(2n+1)k}$, $\frac{n(k-1)}{(2n+1)k}$ for $S_1 - T_1$, $S_2 - T_2$ and $S_3 - T_3$ respectively. This means that as $n$ and $k$ become arbitrarily large, a throughput close to $\frac{1}{2}$ can be achieved for every source-destination pair. However, in this section we show that if $X_i'(0)$, for some $i = i_1$, cannot be recovered then, $X_{i_1}'(p)$ is not recoverable for any $p$. Conversely, we also show that if $X_i'(0)$, for all $i$, can be recovered then $X_i'(p)$ is recoverable for all $p$ and $i$.

*Definition 2:* PBNA in 3-S 3-D MUN-D using Transform Approach and Block Time Varying LEKs is said to be feasible if $X_i'(p)$ can be recovered from $Y_i^{(p \oplus k)}$ for all $i$, for all $p \neq 0$, and for every $n > 1$.

Henceforth, PBNA in 3-S 3-D MUN-D using Transform Approach and Block Time Varying LEKs shall be simply referred to as PBNA in 3-S 3-D MUN-D. We now head towards proving that the reduced feasibility conditions of Meng et al. for feasibility of PBNA in 3-S 3-D I-MUN are also necessary and sufficient for PBNA in 3-S 3-D MUN-D.

PBNA in 3-S 3-D MUN-D is feasible iff there exists a choice of $(n+1) \times n$ matrices $A^{(p)}$ and $B^{(p)}$, $V_1^{(p)}$, and a $n \times n$ matrix $C^{(p)}$ ($0 \leq p \leq k-1$), all with entries from $\mathbb{F}_{2^m}$, such that [12]

$$\det[V_1^{(p)} \quad M_{11}^{p\,-1} M_{21}^p M_{23}^{p\,-1} M_{13}^p V_1^{(p)} A^{(p)}] \neq 0,$$
$$\det[M_{12}^{p\,-1} M_{22}^p M_{23}^{p\,-1} M_{13}^p V_1^{(p)} A^{(p)} \quad V_1^{(p)}] \neq 0,$$
$$\det[M_{13}^{p\,-1} M_{33}^p M_{32}^{p\,-1} M_{12}^p V_1^{(p)} B^{(p)} \quad V_1^{(p)}] \neq 0,$$
$$U^{(p)} V_1^{(p)} AC = V_1^{(p)} B.$$

where, $U^{(p)} = M_{12}^{p\,-1} M_{32}^p M_{31}^{p\,-1} M_{21}^p M_{23}^{p\,-1} M_{13}^p$. The above conditions are obtained from the network alignment conditions in (10). For $0 \leq p \leq k-1$, define

$$\eta(p) = \frac{M_{21}(\underline{\varepsilon}, \alpha^p) M_{32}(\underline{\varepsilon}, \alpha^p) M_{13}(\underline{\varepsilon}, \alpha^p)}{M_{31}(\underline{\varepsilon}, \alpha^p) M_{23}(\underline{\varepsilon}, \alpha^p) M_{12}(\underline{\varepsilon}, \alpha^p)},$$

$$b_1(p) = \frac{M_{21}(\underline{\varepsilon}, \alpha^p) M_{13}(\underline{\varepsilon}, \alpha^p)}{M_{11}(\underline{\varepsilon}, \alpha^p) M_{23}(\underline{\varepsilon}, \alpha^p)}, \quad b_2(p) = \frac{M_{22}(\underline{\varepsilon}, \alpha^p) M_{13}(\underline{\varepsilon}, \alpha^p)}{M_{12}(\underline{\varepsilon}, \alpha^p) M_{23}(\underline{\varepsilon}, \alpha^p)},$$

$$b_3(p) = \frac{M_{33}(\underline{\varepsilon}, \alpha^p) M_{12}(\underline{\varepsilon}, \alpha^p)}{M_{13}(\underline{\varepsilon}, \alpha^p) M_{32}(\underline{\varepsilon}, \alpha^p)}.$$

As in [11], we shall consider the two cases of $\eta(0)$ not being a constant and a constant, separately.

**Case 1:** $\eta(0)$ is a not a constant. The choice of pre-coding matrices are similar to that in [8] [11].

$$V_1^{(p)} = [W \quad U^{(p)}W \quad U^{(p)2}W \quad \cdots \quad U^{(p)n}W], \quad (11)$$
$$V_2^{(p)} = [R^{(p)}W \quad R^{(p)}U^{(p)}W \quad R^{(p)}U^{(p)2}W \quad \cdots \quad R^{(p)}U^{(p)n-1}W],$$
$$V_3^{(p)} = [S^{(p)}U^{(p)}W \quad S^{(p)}U^{(p)2}W \quad \cdots \quad S^{(p)}U^{(p)n}W].$$

where, $R = M_{13}^p {M_{23}^p}^{-1}$, $S = M_{12}^p {M_{32}^p}^{-1}$ ($0 \leq p \leq k-1$), and $W = [1\ 1\ \cdots\ 1]^T$ (all ones vector of size $(2n+1) \times 1$).

*Lemma 1:* PBNA in 3-S 3-D MUN-D is feasible iff, for $1 \leq p \leq k-1$ and for all $i$,
$$b_i(p) \notin \mathcal{S}_n^{(p)} = \left\{ \frac{f(\eta(p))}{g(\eta(p))} \,\bigg|\, f(x), g(x) \in \mathbb{F}_{2^m}[x], f(x)g(x) \neq 0, \right.$$
$$\left. \gcd(f(x), g(x)) = 1,\ \deg(f) \leq n,\ \deg(g) \leq n-1 \right\}.$$
for any $n > 1$.

*Proof:* Proof for sufficiency, with the choice of precoding matrices as given in (11), is similar as that for instantaneous network ($p = 0$) as in [13]. Proof for necessity, taking into account other possible choices of precoding matrices satisfying (10), is the same as that for $p = 0$ case as in [11]. ∎

The following theorem of Meng et al. gives the reduced feasibility conditions for 3-S 3-D I-MUN.

*Theorem 2 ( [11] (Reduced Feasibility Conditions)):* $X_i'(0)$ can be recovered from $Y_i^{(0 \oplus k)}$, for all $i$, iff
$$b_i(0) \notin \mathcal{S}^{(0)} = \left\{ 1, \eta(0), \eta(0) + 1, \frac{\eta(0)}{\eta(0)+1} \right\}. \quad (12)$$

The following theorem shows that $b_i(p) \notin \mathcal{S}_n^{(p)}$ iff $b_i(0) \notin \mathcal{S}^{(0)}$.

*Theorem 3:* When $\eta(0)$ is not a constant, $X_i'(p)$ can be recovered from $Y_i^{(p \oplus k)}$, for all $p$, iff $X_i'(0)$ can be recovered from $Y_i^{(0 \oplus k)}$.

*Proof:* First, we shall reduce the infinite set $\cup_{n=1}^\infty \mathcal{S}_n^{(p)}$ to the form similar to (12). We shall prove that if, for $p \neq 0$,
$$b_i(p) \notin \mathcal{S}^{(p)} = \left\{ 1, \eta(p), \eta(p) + 1, \frac{\eta(p)}{\eta(p)+1} \right\} \quad (13)$$

then, $b_i(p) \notin \cup_{n=1}^\infty \mathcal{S}_n^{(p)}$ as well where, $\mathcal{S}_n^{(p)}$ is defined in Lemma 1. The linearization and square-term property [11] hold good and following exactly the same steps as in [11], $\cup_{n=1}^\infty \mathcal{S}_n^{(p)}$ can be reduced to the set of the form $\left\{ \frac{h_0 + h_1 \eta(p)}{h_2 + h_3 \eta(p)} \right\}$ where, $h_0, h_1, h_2, h_3 \in \mathbb{F}_{2^m}$, $(h_0, h_1) \neq (0, 0)$ and $(h_2, h_3) \neq (0, 0)$. We shall now prove that if $b_i(p) \notin \mathcal{S}^{(p)}$ then
$$b_i(p) \notin \left\{ \frac{h_1 \eta(p)}{h_2 + h_3 \eta(p)}\ (h_0 h_1 h_2 \neq 0),\ h_0 + h_1 \eta(p)\ (h_0 h_1 \neq 0) \right\}.$$

The proof for rest of the cases of the set $\left\{ \frac{h_0 + h_1 \eta(p)}{h_2 + h_3 \eta(p)} \right\}$ is exactly the same as in [11]. We shall prove that $b_i(p) \notin \frac{h_1 \eta(p)}{h_2 + h_3 \eta(p)}$ ($h_1 h_2 h_3 \neq 0$) and the other case can be proved similarly. Let $b_i(p) \notin \mathcal{S}^{(p)}$ and suppose that $b_i(p) = \frac{h_1 \eta(p)}{h_2 + h_3 \eta(p)}$ for some $h_1, h_2, h_3 \in \mathbb{F}_{2^m} \setminus \{0\}$. Consider $i = 1$ (the rest can be proved in the same way). Substituting for $b_1(p)$ and $\eta(p)$, we have the equality as given in (14) (at the top of the next page) where, $\chi_{ij}$ represents the set of all paths from $S_i$ to $T_j$, $G(P_{ij})$ represents the product of LEKs along the path $P_{ij}$ (i.e., the path gain) and $d_{P_{ij}}$ represent the integer delay along the path $P_{ij}$. We refer to $G(P_{11})G(P_{23})G(P_{32})$ as *monomial* and $G(P_{11})G(P_{23})G(P_{32})\alpha^{p(d_{P_{11}} + d_{P_{23}} + d_{P_{32}})}$ as *monomial term*. Since min-cut between $S_i - T_j$ is non-zero for all $(i, j)$, by Menger's theorem, there exists at least one non-zero monomial term on the L.H.S of (14). Clearly, the monomial terms of the L.H.S cannot cancel among themselves as different monomial terms contain the product of path gains of different path tuples.

Now, consider the following two cases separately - (a) every monomial term on the L.H.S cancels with one monomial term on the R.H.S, (b) Some monomial term on the L.H.S cancels with the sum of one monomial term from $Sum_1$ and another monomial term from $Sum_2$.

(a) Every monomial term on the L.H.S cancels with a monomial term either from $Sum_1$ or $Sum_2$. Suppose that $G(P_{11})G(P_{23})G(P_{32})\alpha^{p(d_{P_{11}} + d_{P_{23}} + d_{P_{32}})}$ cancels with $\frac{h_2}{h_1} G(P_{12})G(P_{23})G(P_{31})\alpha^{p(d_{P_{12}} + d_{P_{23}} + d_{P_{31}})}$. Clearly, the LEKs occurring in $G(P_{11})G(P_{23})G(P_{32})$ has to be the same as that in $G(P_{12})G(P_{23})G(P_{31})$. Therefore, it means that every edge covered by the path tuple $(P_{11}, P_{23}, P_{32})$ are also covered by edges in the path tuple $(P_{12}, P_{23}, P_{31})$ including multiplicities. Hence, $d_{P_{11}} + d_{P_{23}} + d_{P_{32}} = d_{P_{12}} + d_{P_{23}} + d_{P_{31}}$. So, $\frac{h_2}{h_1} = 1$. By Menger's theorem, there exists at least one non-zero monomial term in $Sum_2$. A monomial term in $Sum_2$ has to cancel with a monomial term in L.H.S or with a monomial term in $Sum_1$ or with a difference between two monomial terms, one each from L.H.S and $Sum_1$. The last possibility is similar to the one treated in case (b). If a monomial term in $Sum_2$ cancels with a term in L.H.S then, $\frac{h_3}{h_1} = 1$. But, $\frac{h_2}{h_1}$ is also 1. This contradicts the hypothesis that $b_i(p) \notin \mathcal{S}^{(p)}$. If a monomial term in $Sum_2$ cancels with a monomial term in $Sum_1$, then $\frac{h_3}{h_1} = \frac{h_2}{h_1}$. But $\frac{h_2}{h_1} = 1$ which means that $\frac{h_3}{h_1} = 1$. This again contradicts $b_i(p) \notin \mathcal{S}^{(p)}$. Similarly we can prove that if $G(P_{11})G(P_{23})G(P_{32})\alpha^{p(d_{P_{11}} + d_{P_{23}} + d_{P_{32}})}$ cancels with $\frac{h_3}{h_1} G(P_{21})G(P_{32})G(P_{13})\alpha^{p(d_{P_{21}} + d_{P_{32}} + d_{P_{13}})}$, it leads to contradiction of $b_i(p) \notin \mathcal{S}^{(p)}$.

(b) Some monomial term on the L.H.S cancels with the sum of one monomial term from $Sum_1$ and another monomial term from $Sum_2$. Note that a monomial term on the L.H.S cannot be the sum of two monomial terms, both from $Sum_1$ or $Sum_2$, as different monomial terms of $Sum_1$ contain product of path gains of different path tuples. If at all a monomial term on the L.H.S cancels with the sum of two monomial terms, one has to be from $Sum_1$ and the other from $Sum_2$. Let $G(P_{11})G(P_{23})G(P_{32})\alpha^{p(d_{P_{11}} + d_{P_{23}} + d_{P_{32}})}$ cancel with $\frac{h_2}{h_1} G(P_{12})G(P_{23})G(P_{31})\alpha^{p(d_{P_{12}} + d_{P_{23}} + d_{P_{31}})} + \frac{h_3}{h_1} G(P_{21})G(P_{32})G(P_{13})\alpha^{p(d_{P_{21}} + d_{P_{32}} + d_{P_{13}})}$. Now, the monomials $G(P_{12})G(P_{23})G(P_{31})$ and $G(P_{21})G(P_{32})G(P_{13})$ have to contain the same variables as that in $G(P_{11})G(P_{23})G(P_{32})$. So, $d_{P_{11}} + d_{P_{23}} + d_{P_{32}} = d_{P_{12}} + d_{P_{23}} + d_{P_{31}} = d_{P_{21}} + d_{P_{32}} + d_{P_{13}}$. Now, note that such a cancellation can happen iff $\frac{h_2 + h_3}{h_1} = 1$. If there exists a monomial term on the L.H.S that cancels with a monomial term from $Sum_1$ ($Sum_2$), then $\frac{h_2}{h_1} = 1$ $\left( \frac{h_3}{h_1} = 1 \right)$ which means $h_3 = 0$ ($h_2 = 0$). This contradicts $b_1(p) \notin \mathcal{S}^{(p)}$. So, let *every* monomial term on the L.H.S cancel with the sum of one monomial term from $Sum_1$ and another monomial term from $Sum_2$. After these cancellations if there are no monomial terms left over on the R.H.S of (14) then clearly, the equality in (14) also holds good for $\left( \frac{h_2}{h_1}, \frac{h_3}{h_1} \right) = (1, 0)$ or $\left( \frac{h_2}{h_1}, \frac{h_3}{h_1} \right) = (0, 1)$ which contradicts $b_i(p) \notin \mathcal{S}^{(p)}$. If there are monomial terms left over in $Sum_1$

$$\frac{h_1\eta(p)}{h_2+h_3\eta(p)}=b_1(p)\Rightarrow M_{11}(\underline{\varepsilon},\alpha^p)M_{23}(\underline{\varepsilon},\alpha^p)M_{32}(\underline{\varepsilon},\alpha^p)=\frac{h_2}{h_1}M_{12}(\underline{\varepsilon},\alpha^p)M_{23}(\underline{\varepsilon},\alpha^p)M_{31}(\underline{\varepsilon},\alpha^p)+\frac{h_3}{h_1}M_{21}(\underline{\varepsilon},\alpha^p)M_{32}(\underline{\varepsilon},\alpha^p)M_{13}(\underline{\varepsilon},\alpha^p)$$

$$\Rightarrow \sum_{P_{11}\in\chi_{11},P_{23}\in\chi_{23},P_{32}\in\chi_{32}} G(P_{11})G(P_{23})G(P_{32})\alpha^{p(d_{P_{11}}+d_{P_{23}}+d_{P_{32}})} = \frac{h_2}{h_1}\underbrace{\sum_{P_{12}\in\chi_{12},P_{23}\in\chi_{23},P_{31}\in\chi_{31}} G(P_{12})G(P_{23})G(P_{31})\alpha^{p(d_{P_{12}}+d_{P_{23}}+d_{P_{31}})}}_{Sum_1} \quad (14)$$

$$+\frac{h_3}{h_1}\underbrace{\sum_{P_{21}\in\chi_{21},P_{32}\in\chi_{32},P_{13}\in\chi_{13}} G(P_{21})G(P_{32})G(P_{13})\alpha^{p(d_{P_{21}}+d_{P_{32}}+d_{P_{13}})}}_{Sum_2}$$

$$\sum_{P_{11}\in\chi_{11},P_{23}\in\chi_{23},P_{32}\in\chi_{32}} G(P_{11})G(P_{23})G(P_{32}) = \underbrace{\sum_{P_{12}\in\chi_{12},P_{23}\in\chi_{23},P_{31}\in\chi_{31}} G(P_{12})G(P_{23})G(P_{31})}_{Sum'_1}+\underbrace{\sum_{P_{21}\in\chi_{21},P_{32}\in\chi_{32},P_{13}\in\chi_{13}} G(P_{21})G(P_{32})G(P_{13})}_{Sum'_2} \quad (15)$$

and $Sum_2$ after the cancellation of monomial terms of L.H.S of (14), then every remaining monomial term of $Sum_1$ has to cancel with that of $Sum_2$ which means that $h_2 = h_3$. But, $\frac{h_2+h_3}{h_1} = 1$. Since the field of operation is of characteristic 2 and $h_1 \neq 0$, $\frac{h_2+h_3}{h_1} = 1$ cannot be satisfied. Hence, case $(b)$ also leads to contradiction of $b_1(p) \notin \mathcal{S}^{(p)}$.

To prove the theorem, we now need to show that $b_i(p) \in \mathcal{S}^{(p)}$ ($p \neq 0$) iff $b_i(0) \in \mathcal{S}^{(0)}$. We shall assume $i = 1$ and the proof for the rest are similar.

*If Part:* Suppose $b_1(0) = \frac{\eta(0)}{1+\eta(0)}$. Substituting for $b_1(0)$ and $\eta(0)$, we get (15). As in the earlier part of the proof, the terms on the L.H.S of (15) cannot cancel among themselves. So, every term in the L.H.S has to cancel with a term in $Sum'_1$ or $Sum'_2$. If the term in the L.H.S cancels with a term in $Sum'_1$ then $d_{P_{11}} + d_{P_{23}} + d_{P_{32}} = d_{P_{12}} + d_{P_{23}} + d_{P_{31}}$; if the cancellation is with a term in $Sum'_2$ then $d_{P_{11}} + d_{P_{23}} + d_{P_{32}} = d_{P_{21}} + d_{P_{32}} + d_{P_{13}}$. The remaining un-canceled terms in $Sum'_1$ has to cancel with the un-canceled terms in $Sum'_2$. For these terms, $d_{P_{12}} + d_{P_{23}} + d_{P_{31}} = d_{P_{21}} + d_{P_{32}} + d_{P_{13}}$. Hence, (14) is satisfied with $\frac{h_2}{h_1} = \frac{h_3}{h_1} = 1$. Therefore, $b_1(p) = \frac{\eta(p)}{1+\eta(p)}$, $\forall\ p \neq 0$. Similarly it can be proved that if $b_i(0)$ belongs to any other element of $\mathcal{S}^{(0)}$ then, $b_i(p) \in \mathcal{S}^{(p)}$, $\forall\ p \neq 0$.

*Only If Part:* Assume that $b_1(p_1) = \frac{\eta(p_1)}{1+\eta(p_1)}$ for some $p_1 \neq 0$. Following the same steps as in the "If Part" regarding cancellation of terms one can prove that $b_1(p) = \frac{\eta(p)}{1+\eta(p)}$, $\forall\ p \neq p_1$ which includes $p = 0$.

Hence, $b_i(p) \notin \mathcal{S}^{(p)}$ ($p \neq 0$) iff $b_i(0) \notin \mathcal{S}^{(0)}$. ∎

In brief, the above theorem proves that the reduced feasibility conditions of Meng et al. for feasibility of PBNA in 3-S 3-D I-MUN are also necessary and sufficient for feasibility of PBNA in 3-S 3-D MUN-D when $\eta(0)$ is not a constant.

**Case 2:** $\eta(0)$ is a constant. When $\eta(0)$ is a constant, Theorem 1 of [11] states that $X'_i(0)$ can be recovered from $Y_i^{(0\oplus k)}$ iff $b_i(0)$ is not a constant for each $i \in \{1, 2, 3\}$. Similar to Theorem 1 of [11] we have the following lemma.

*Lemma 2:* PBNA in 3-S 3-D MUN-D is feasible iff $b_i(p)$ is not a constant for each $i \in \{1, 2, 3\}$ and for $1 \leq p \leq k-1$.

*Proof:* Proof is the same as for $p = 0$ case in [11]. ∎

The following proposition in combination with Theorem 1 of [11] and Lemma 2 shows that PBNA in a 3-S 3-D MUN-D is feasible iff PBNA in the 3-S 3-D I-MUN is feasible.

*Proposition 1:* $b_i(p)$ is a constant iff $b_i(0)$ is a constant.

*Proof:* The proof follows using the same arguments as in "If Part" and "Only If Part" in the proof of Theorem 3. ∎

The feasibility conditions for PBNA in 3-S 3-D MUN-D for the case of zero min-cut between $S_i - T_j$ for some $(i, j)$ are also the same as that for 3-S 3-D I-MUN as given in [11].

## V. CONCLUSION

A new PBNA scheme for 3-S 3-D MUN-D is proposed which is different from PBNA with time-invariant LEKs and time-varying LEKs [12] where, the independent symbols are precoded within a single block of data to be transmitted after addition of CP and pre-multiplication by $F$. In the proposed PBNA scheme, the independent symbols are precoded across multiple blocks of data which are demarcated by separate CPs. We showed that the proposed PBNA scheme inherits the reduced feasibility conditions of Meng et al. The motivation for the new scheme was that the feasibility of PBNA with time-invariant LEKs and time-varying LEKs [12] could not be easily checked. However, a caveat in the proposed PBNA scheme is that the decoding delay is higher compared to that for PBNA with time-invariant LEKs and time-varying LEKs.

Using the proof technique of Theorem 3, it can shown that the feasibility conditions for PBNA in 3-S 3-D I-MUN are also necessary conditions for feasibility of PBNA with time-invariant LEKs in 3-S 3-D MUN-D. However, sufficiency of the conditions remain open. Further, necessary and sufficient conditions for feasibility of PBNA with time-varying LEKs in 3-S 3-D MUN-D are known only for a given value of symbol extensions and it is not known if PBNA using time-varying LEKs is feasible when PBNA using transform approach and block time varying LEKs is not feasible.